\documentclass[aps,amsmath,amssymb,prl,reprint,floatfix,superscriptaddress,longbibliography]{revtex4-2}
\usepackage[T1]{fontenc}
\usepackage{graphicx}
\usepackage{amsmath}
\usepackage{amssymb}
\usepackage{bm}
\usepackage{subcaption}
\usepackage[colorlinks=true, allcolors=blue]{hyperref}
\usepackage{times}
\usepackage[normalem]{ulem}

\begin{document}

\title{Superconductivity and Polar Metallicity in Atomically Thin Nitride Superlattices}

\author{Huma Nawaz}  
\affiliation{Texas Center for Superconductivity and Department of Physics, University of Houston, Houston, TX 77204, USA}

\author{Dawei Wang} 
\affiliation{Texas Center for Superconductivity and Department of Physics, University of Houston, Houston, TX 77204, USA}

\author{C. S. Ting} 
\email{cting2@Central.UH.EDU}
\affiliation{Texas Center for Superconductivity and Department of Physics, University of Houston, Houston, TX 77204, USA}

\author{Laurent Bellaiche} 
\email{laurent@uark.edu}
\affiliation{Smart Ferroic Materials Center, Department of Physics and Institute for Nanoscience and Engineering, University of Arkansas, Fayetteville, Arkansas 72701, USA} 
\affiliation{Department of Materials Science and Engineering, Tel Aviv University, Ramat Aviv, Tel Aviv 6997801, Israel}

\date{\today}

\begin{abstract}
The integration of polar and superconducting materials in the atomically thin
limit typically results in a competition that suppresses the superconducting
order. Here, we challenge this conventional view by employing first-principles calculations to investigate 1:1 superlattices 
composed of alternating atomically thin layers of TiN/AlN and $\delta$-NbN/AlN.
Contrary to the expectation that dense insulating barriers would quench
superconductivity, we find that the polar interface can coexist with a
superconducting state. We observe that the transition temperature ($T_c$) of
TiN is enhanced by nearly 2 K over its bulk value, while $\delta$-NbN maintains a relatively high $T_c$ exceeding 20 K. We demonstrate that this resilience arises from
the formation of a polar metal, where the strong internal electric fields of
the AlN layers are partially screened by a charge redistribution within the
nitride layers. This screening mechanism, along with the geometric confinement,
creates a quasi-two-dimensional electronic environment, establishing these superlattices as a novel
platform for exploring the coexistence of structural 
inversion-symmetry breaking and superconductivity in 
the two-dimensional limit.
\end{abstract}

\maketitle

Transition metal nitrides (TMNs), such as titanium nitride (TiN) and niobium
nitride (NbN), stand as cornerstones of modern quantum technology. Their unique
combination of metallic conductivity, chemical inertness, and robust
superconductivity makes them indispensable for a broad spectrum of
applications, ranging from plasmonic waveguides to superconducting quantum
circuits \cite{shalaev_plasmonics_2007}. NbN, in particular, boasts a
relatively high superconducting transition temperature ($T_{c}\approx17$ K)
\cite{williams1967superconductivity}, making it a critical material for
superconducting nanowire single-photon detectors \cite{vodolazov2020timing,
esmaeil2021superconducting, iovan2023nbn} and high-quality compact resonators
\cite{frasca2023nbn}. A persistent challenge, however, lies in the degradation
of these desirable properties in the ultrathin film regime, where
substrate-induced disorder, strain, and defects often suppress
superconductivity \cite{horne2024optimization}.

To overcome these limitations, significant experimental effort has focused on
engineering epitaxial heterostructures, particularly by integrating NbN with
aluminum nitride (AlN) \cite{iovan2023nbn, yang2023cmos, kobayashi2020coherent,
goldsmith2018influence}. The wurtzite AlN (0001) surface provides a nearly
perfect lattice-matched template for the growth of (111)-oriented rocksalt NbN,
with a lattice mismatch as low as $+0.61\%$ \cite{mohanbabu2021design} (for AlN
and TiN the lattice mismatch is $-3.6\%$).  This epitaxial relationship
drastically reduces crystalline defects, leading to a remarkable enhancement of
the superconducting properties of ultrathin NbN films, bringing their $T_{c}$
closer to the bulk limit \cite{brisson2025enhancement}. Consequently, the
AlN/NbN interface has become a foundational platform for state-of-the-art
superconducting devices.

While AlN is traditionally viewed merely as a passive structural buffer that
improves crystallinity \cite{brisson2025enhancement}, the physical consequences
of pushing this heterostructure to its ultimate quantum limit, which results in
an atomically alternating superlattice, remain largely unexplored. In such a
1:1 geometry, where the superconductor is densely interwoven with polar
insulating layers, one might intuitively expect a severe suppression of
superconductivity due to the reduction in charge carrier density and the
disruption of long-range phase coherence. In this study, we report a strikingly
anomalous behavior that contradicts this intuition.  In fact, superconductivity
in 1:1 TiN/AlN and $\delta$-NbN/AlN superlattices is not only resilient but can
be enhanced compared to the bulk parent compounds. Using first-principles
density functional theory (DFT), we show that this phenomenon arises from the
formation of an intrinsic polar metal state \cite{zhou2020review,
bhowal2023polar}. In addition to the geometric confinement, the presence of
polar AlN layers help inducing a charge redistribution that preserves the
metallic density of states (DOS) and generates flat bands near the Fermi level.

\begin{figure}[t]
\includegraphics[scale=0.35]{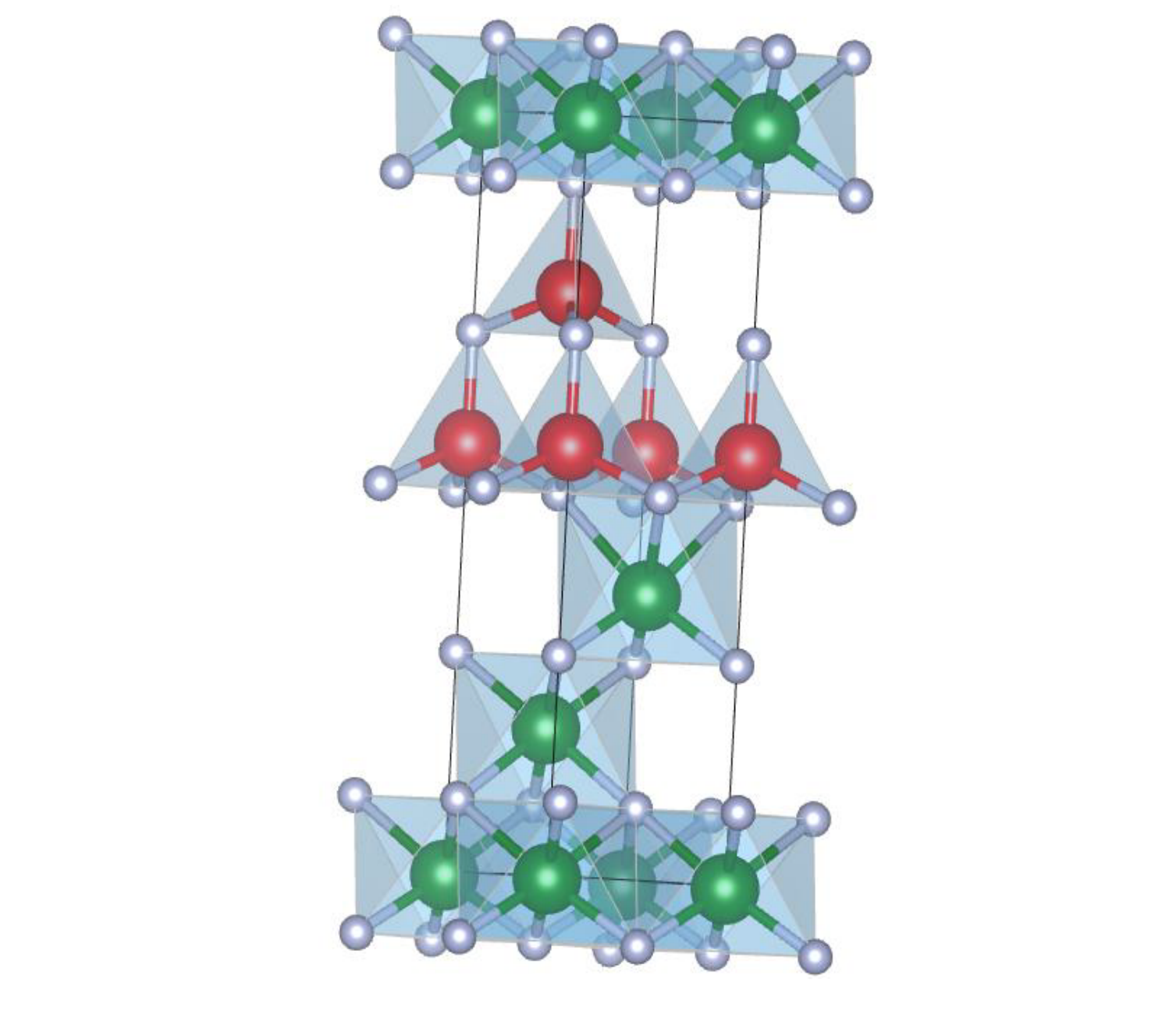}
\caption{Relaxed crystal structure of the 1:1 Ti($\delta$-Nb)N/AlN
Ti/Nb atoms are shown in green, Al in red, and N in grey.}
\label{fig:Superlattices}
\end{figure}

We performed our calculations using the Quantum ESPRESSO (QE) package
\cite{giannozzi2017advanced} with the Perdew-Burke-Ernzerhof (PBE)
exchange-correlation functional \cite{perdew1996generalized} and Optimized
Norm-conserving Vanderbilt Pseudopotentials (ONCVPSP v0.5) from Pseudo dojo repository \cite{hamann2013optimized, van2018pseudodojo} with a
kinetic energy cutoff of 80 Ry. The superlattices were constructed by stacking
alternating layers along the [111] direction (Fig.\,\ref{fig:Superlattices}) and
fully relaxed to a force convergence threshold of $10^{-6}$ a.u., resulting in a
structure with $P3m1$ symmetry. We note that the parent structures for this
construction are rocksalt TiN ($a=4.24$\,{\AA}), rocksalt $\delta$-NbN
($a=4.425$\,{\AA}), and wurtzite AlN ($a=3.11$\,{\AA} and $c=4.98$\,{\AA})
\cite{strite1992gan}. We also note that NbN can exist in four polymorphic
structures: cubic ($\delta$-NbN), $\epsilon$-NbN \cite{terao1965structure},
hexagonal \cite{terao1971new}, and tungsten carbide type
\cite{babu2019electron}; here we focus on the cubic phase relevant to
high-$T_c$ applications. The electronic band structure and DOS were calculated using $20\times20\times20$ $\bm{k}$ grid density, and phonon spectra were
investigated using a $10\times10\times10$/$15\times15\times6$ Monkhorst-Pack $\bm{k}$-point grid and a $5\times5\times5$/$5\times5\times3$ $\bm{q}$-point grid for TiN/AlN and $\delta$-NbN/AlN superlattice, respectively for density functional perturbation theory (DFPT).  We also apply a Methfessel-Paxton smearing width ($\sigma$) of
0.02 Ry and a Gaussian smearing (to get the smooth DOS)  $\sigma$=0.12 Ry for the TiN/AlN and $\delta$-NbN/AlN superlattices,
respectively. The large value for the latter is necessary to ensure
structural stability \cite{babu2019electron, isiaho2021structural, bi2025rise, kogler2025vacancy}. For $\delta$-NbN/AlN superlattice, we decoupled the smearing parameter used for phonon dynamical matrices ($\sigma_{ph}=0.12 Ry$) and the smearing parameter used for electronic Fermi-surface integrations ($\sigma_{el}=0.02 Ry$) as has also been reported in Ref. \cite{kogler2025vacancy}. Such decoupling of smearing helps us avoid overestimating $T_c$ obtained at effective smearing as can be seen in Tab.\,\ref{tab:summary}. We have tested different $\sigma$ values and found that 0.12 Ry indeed provides a stable structure for $\delta$-NbN/AlN superlattice. The $T_c$ is estimated using the
McMillan-Allen-Dynes formula \cite{allen1975transition,
mcmillan1968transition}, which depends on the effective screened Coulomb
repulsion, $\mu^*$. The latter is an empirical parameter, typically ranging from 0.05
to 0.25 \cite{gao2017prediction, bo2021high, iyakutti2006electronic, gao2025,
lu2020phonon, penev2016can}.

\begin{figure}[t]
\includegraphics[scale=0.5]{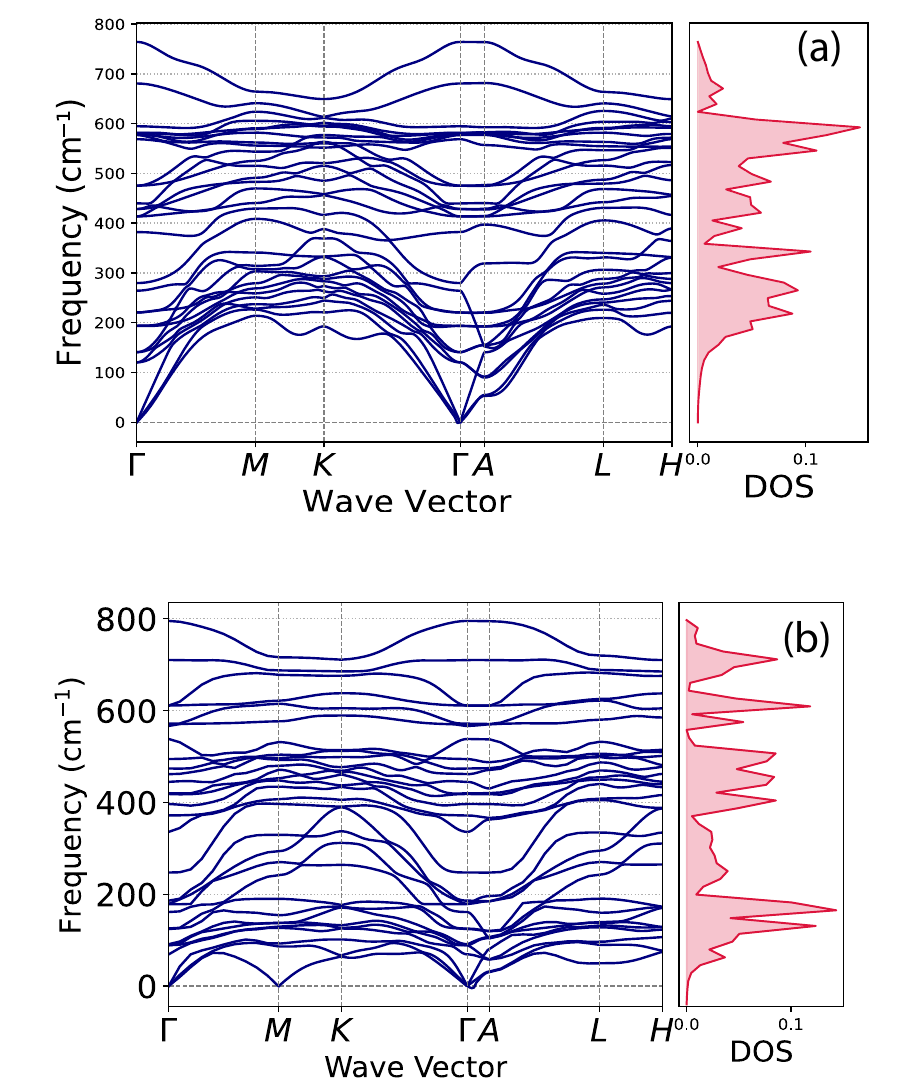} 
\caption{Phonon dispersion relations and phonon DOS for (a) TiN/AlN and (b) $\delta$-NbN/AlN superlattices.}
\label{Fig: TiN/AlN_phonon_dispersion}
\end{figure}

To establish a baseline, we first benchmarked the parent compounds. Our results
for bulk TiN and $\delta$-NbN are in excellent agreement with previous reports
\cite{altintas2013high, babu2019electron}, confirming their metallic nature
driven by Ti-$3d$/N-$2p$ and Nb-$4d$/N-$2p$ hybridization, respectively. For
the superlattices, the calculated phonon dispersions (Fig.\,\ref{Fig:
TiN/AlN_phonon_dispersion}) confirm the dynamical stability of both
superlattices. For the $\delta$-NbN/AlN system, the minor
imaginary frequencies near the $\Gamma$ point (Fig.\,\ref{Fig: TiN/AlN_phonon_dispersion}b) are understood to be
numerical artifacts from convergence challenges rather than structural
instabilities, consistent with previous observations \cite{lu2020phonon}. By testing different k-grid densities ($10\times10\times4$ versus $15\times15\times 6$), we verified the numerical origin of these unphysical frequencies and the convergence of $T_c$ for $\delta$-NbN/AlN system. Notably, the distinct phonon gaps characteristic of bulk TiN (around 410-650
cm$^{-1}$ \cite{altintas2013high}) and $\delta$-NbN (around  200-370 cm$^{-1}$
\cite{babu2019electron}) are closed in the superlattices.  This occurs due to
hybridization with the gapless phonon spectrum of wurtzite AlN
\cite{bungaro2000ab}. We will see in Fig.\,\ref{fig:a2F, lambda} that such
significant changes in the vibrational properties can affect superconductivity: In the TiN/AlN system, the electron-phonon coupling (EPC) constant ($\lambda(\omega)$) is enhanced by high-frequency optical modes, whereas in $\delta$-NbN/AlN, the coupling is dominated by the low-frequency peak around 150 cm$^{-1}$ and the intermediate-frequency (400-600 cm$^{-1}$) modes.

\begin{figure}[t] 
\includegraphics[scale=0.4]{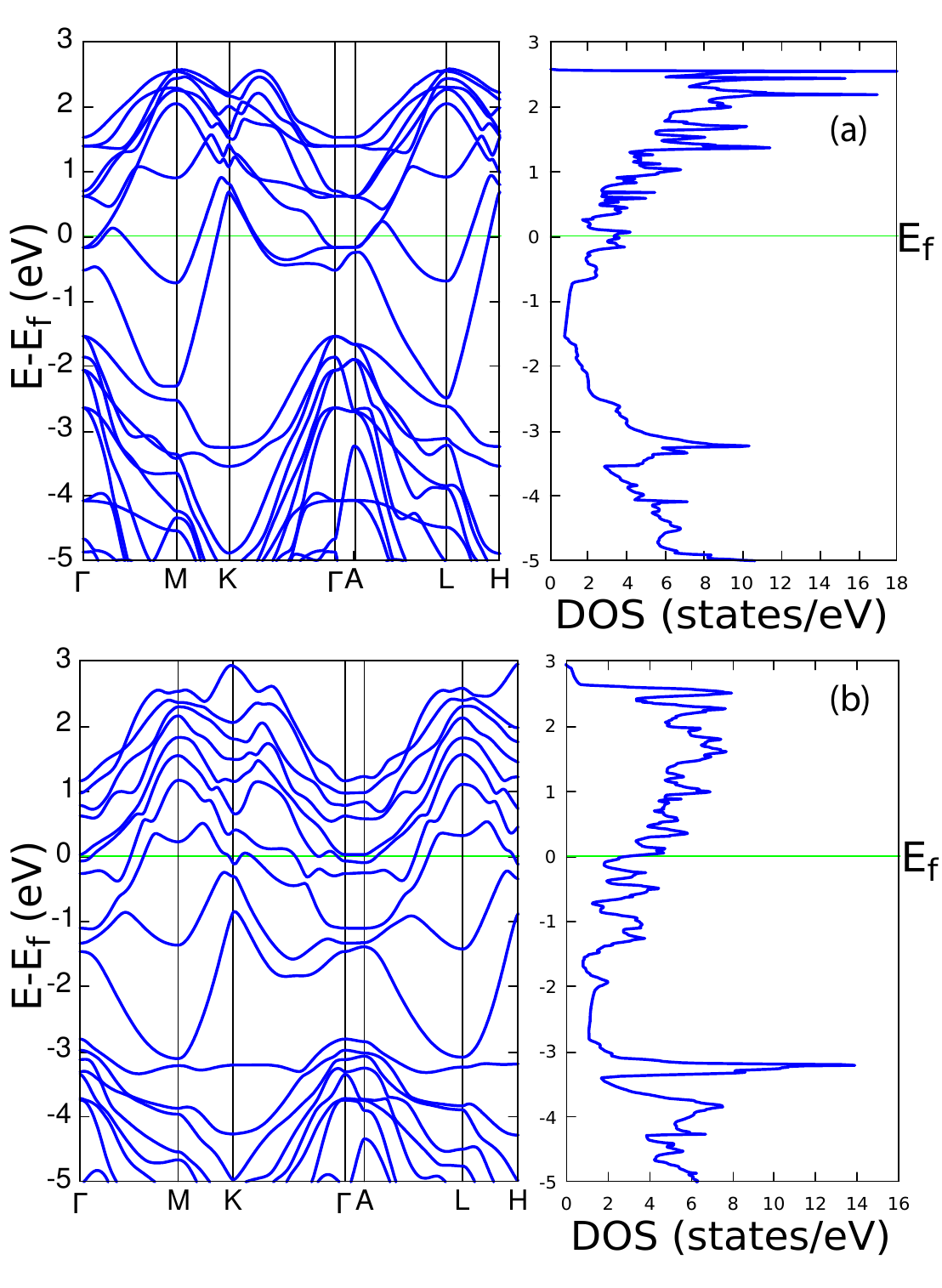} 
\caption{
 Electronic band structures and DOS for (a) TiN/AlN and (b) $\delta$-NbN/AlN.
}
\label{Fig: TiN/AlN} 
\end{figure}

To understand the electronic origin of the superconductivity, we examine the
band structures presented in Fig.\,\ref{Fig: TiN/AlN}. A striking feature common
to both systems is the appearance of a nearly flat band located at the Fermi
level along the $\Gamma$-A direction. Since the $\Gamma$-A path corresponds to
the out-of-plane axis ($k_z$), the minimal dispersion of this band implies a
vanishing group velocity \cite{raza2020atio3} and a very large effective mass for charge carriers
along the stacking direction. This signifies strong electron confinement within
the metallic nitride layers, effectively creating a quasi-two-dimensional
electron system. The emergence 
of these flat bands along the $\Gamma$-$A$ (i.e., the stacking direction) is physically 
significant because it collapses the carrier group velocity and may drive a 
sharp accumulation of spectral weight near 
the Fermi level. Within the Migdal-Eliashberg theory, such a 
localized enhancement of the electronic DOS can enhance the electron-phonon coupling 
parameter, $\lambda$, thereby  
stabilizing the $T_c$ despite extreme 
structural confinement in the vertical direction. 

\begin{figure}[h]
\includegraphics[scale=0.4]{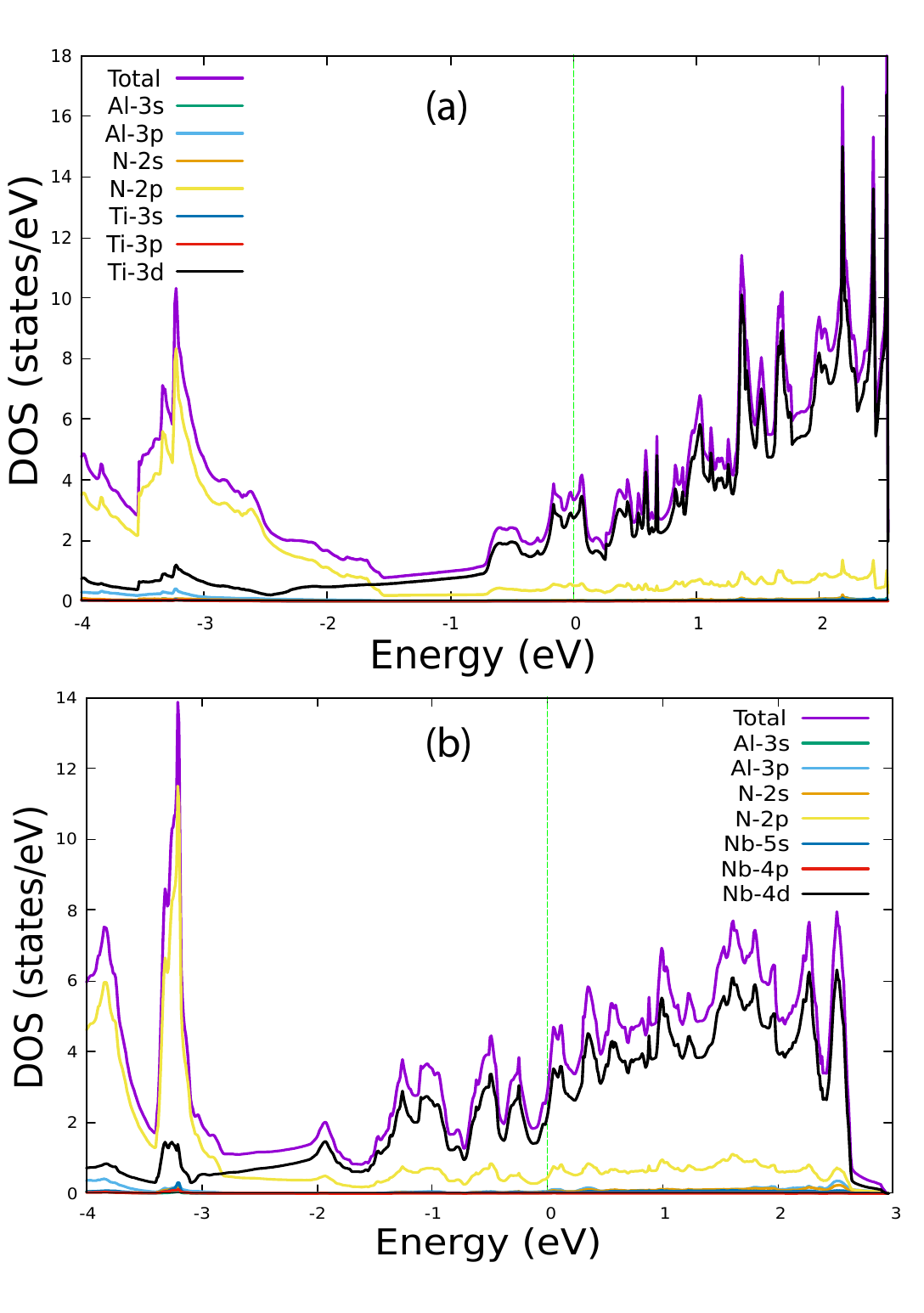}\\
  \caption{Orbital-resolved projected density of states (PDOS) of the (a) TiN/AlN and (b) $\delta$-NbN/AlN superlattice.}
\label{fig: Proj_band}
\end{figure}

As visualized in the PDOS (Fig.\,\ref{fig:
Proj_band}), the flat band in the TiN/AlN and $\delta$-NbN/AlN superlattices is derived
almost exclusively from Ti and Nb-$d$ orbitals. This orbital character 
confirms that the heavy quasiparticles are intrinsic to the superconducting 
TiN and NbN layers, confined by the potential barriers of the polar AlN 
spacers. By engineering this confinement, the superlattice architecture 
essentially creates an ideal flat-band metal, where the DOS is pinned to the 
Fermi level,
thereby amplifying the susceptibility to pairing instability. In addition to
the geometric confinement arising from the superlattice, the strong internal
electric fields from the polar AlN layers break inversion symmetry and induce a
significant charge redistribution, leading to the formation of a
two-dimensional electron gas (2DEG) \cite{yan2018gan} confined at the
interfaces.

\begin{figure}[h]
    \centering
       \includegraphics[scale=0.4]{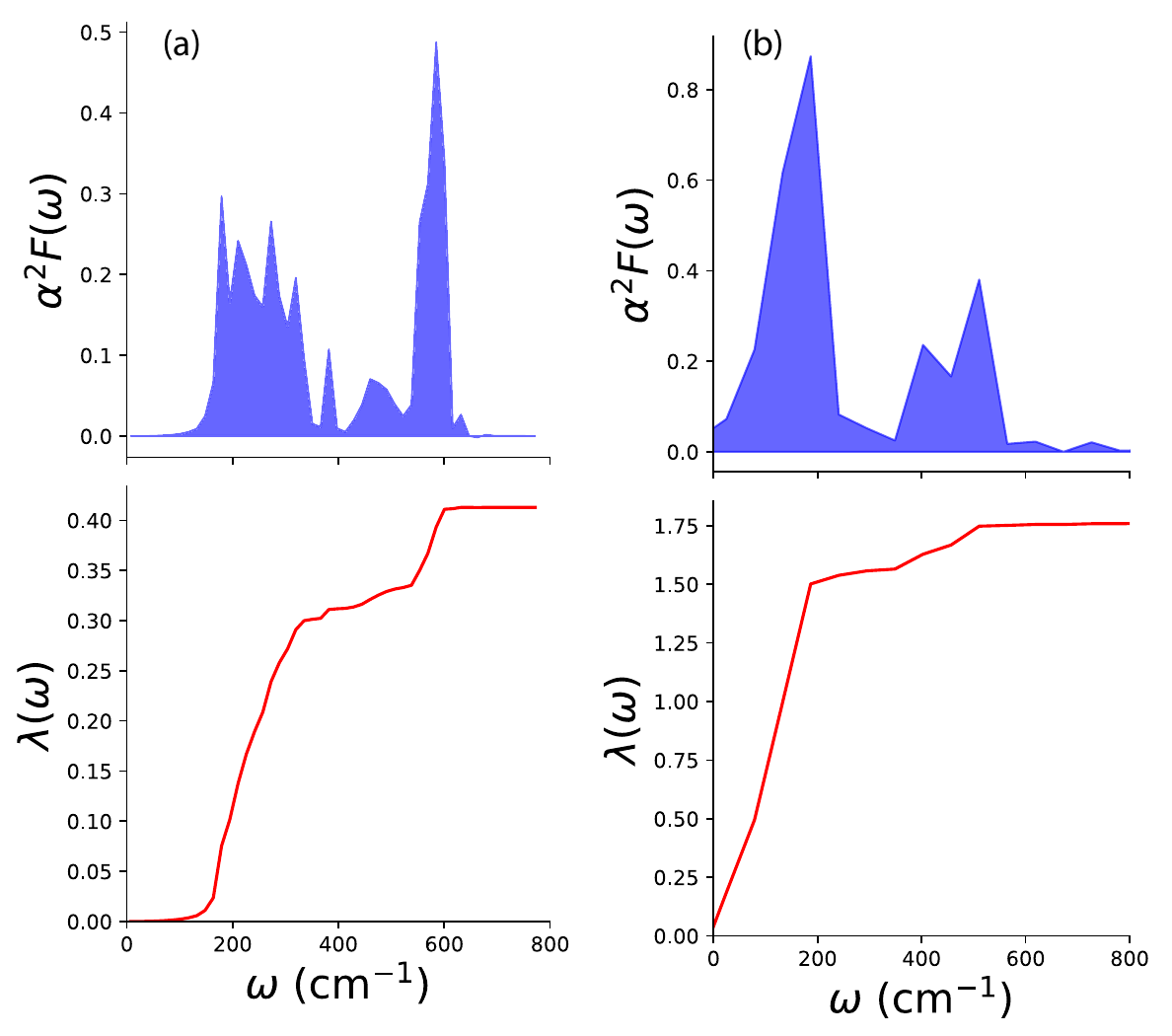}
    \caption{Eliashberg spectral function $\alpha^2F(\omega)$ (shaded blue) and cumulative EPC constant $\lambda(\omega)$ (red line) for (a) TiN/AlN and (b) $\delta$-NbN/AlN (for $\sigma_{el}=0.12$ Ry) superlattice.}
\label{fig:a2F, lambda}
\end{figure}

\begin{table}[htb]
    \centering
    \begin{tabular}{|c|c|c|c|c|}
    \hline
       Structure & $\sigma_{el}$ (Ry) & $\omega_{\rm log}$ (K) & $\lambda$ & $T_c$ (K) \\
       \hline
       TiN          &  0.02 & 613.80 & 0.43 & 4.66   \\
       $\delta$-NbN  & 0.12 & 181.50 & 1.83  & 25.10  \\
       TiN/AlN       & 0.02 & 416.60 & 0.51  &  6.54  \\ 
       $\delta$-NbN/AlN & 0.02 & 188.64  & 0.86 & 10.83\\
       $\delta$-NbN/AlN & 0.12 & 165.70 & 1.75 &   22.22 \\
       \hline
    \end{tabular}
    \caption{
    Calculated superconducting properties for bulk parent compounds and their respective superlattices using $\mu^*=0.09$.}
    \label{tab:summary}
\end{table}

Having obtained the phonon spectra and the band structure, we proceeded to
calculate $T_c$ using $\mu^*=0.09$, as summarized in Tab.\,\ref{tab:summary}. We have found that the EPC constant λ can increase with $\sigma$ for the same system.
The TiN/AlN superlattice exhibits a $T_c$ of 6.5 K, representing a substantial
enhancement of nearly 2 K over the bulk TiN value (4.6 K). To identify the
microscopic mechanism driving this enhancement, we analyze the Eliashberg spectral function estimated by 
$\alpha^2F(\omega)=\frac{1}{N(E_{F})N_k N_q}\sum_{\textbf{k,q},\nu,i,j}| g_{\textbf{k}i, \textbf{k}+\textbf{q}j}^{\textbf{q}\nu}|\delta(\omega-\omega_\textbf{q}\nu)\times\delta(\varepsilon_{\textbf{k}i}-E_F)
\delta(\varepsilon_{\textbf{k}+\textbf{q}j}-E_F)$, here $N_k$ and $N_q$ 
are the total number of k and q points, respectively, 
$\omega_{\textbf{q}\nu}$ is the phonon frequency of the $\nu$th phonon mode at vector \textbf{q}, $N(E_F)$ is the EDOS at the fermi level, $\varepsilon_{\textbf{k}i}$ and $\varepsilon_{\textbf{k}+\textbf{q}j}$ are the Kohn-Sham orbitals at given bands and wave vectors, and $g_{\textbf{k}i, \textbf{k}+\textbf{q}j}^{\textbf{q}\nu}$ is the EPC matrix element. 
The EPC constant $\lambda(\omega)$ and logarithmic average frequency $\omega_{log}(\omega)$ can be determined by $\lambda(\omega)=2\int_0^\omega\frac{\alpha^2 F(\omega^{\prime})}{\omega^{\prime}}d\omega^{\prime}$ \cite{lu2020phonon},
$\alpha^2$$F(\omega)$ and $\lambda(\omega)$
are shown in Fig.\,\ref{fig:a2F, lambda}(a) for TiN/AlN superlattice. The formation of the superlattice introduces new vibrational modes
associated with the lighter AlN layers. The value of $N(E_F)$ increases for the superlattices, as compared to the bulk parent compounds due to the hybridization of orbitals and as shown in Fig.\,\ref{fig: Proj_band}. Crucially, we observe that for TiN/AlN, the $\lambda(\omega)$ receives significant
contributions from high-frequency optical modes around 600 cm$^{-1}$, which is a feature absent in the pure bulk limit. While the introduction of high-frequency modes typically tends to increase $\omega_{\mathrm{log}}$, in the TiN/AlN system, a significant redistribution of spectral weight towards lower frequencies dominates, resulting in a net reduction of $\omega_{\mathrm{log}}$. In the case of TiN/AlN,
the substantial boost in the $\lambda$ (increasing from 0.43
in bulk to 0.51 in the superlattice) completely outweighs this reduction. This
indicates a cooperative effect where the polar phonon modes of the AlN spacer
actively couple to the Ti-$d$ electrons, enhancing the pairing.

This finding is particularly interesting when viewed in the context of
theoretical limits on conventional superconductivity. While Trachenko
\textit{et al.} \cite{trachenko2025upper} demonstrated that the maximum possible $T_c$ is theoretically
achieved by shifting the spectral weight of $\alpha^2F(\omega)$ to the highest
frequencies \cite{trachenko2025upper}, the analysis by Gao \textit{et al.}
highlights that this is often unphysical \cite{gao2025} due to the inherent
trade-off between $\lambda$ and $\omega_{\log}$. Our results for TiN/AlN align
with this complex interplay, where an increase in $\lambda$ outweighs the
decrease in $\omega_{\log}$.

Conversely, for the $\delta$-NbN/AlN superlattice, we calculate a $T_c$ of 22.2
K using $\sigma_{el}=0.12 Ry$. Although this is a slight reduction from our calculated bulk value (25.10 K),
it remains exceptionally high. The spectral analysis in Fig.\,\ref{fig:a2F,
lambda}(b) reveals a fundamentally different behavior compared to the TiN case.
Here, the coupling is dominated almost entirely by low-frequency phonons, which
are primarily associated with the vibrations of the heavier Nb atoms. Unlike
the TiN system, the high-frequency modes above 600 cm$^{-1}$ do not couple
efficiently to the electronic states. Consequently, both $\lambda$ and $\omega_{\log}$ experience a modest reduction compared to bulk
$\delta$-NbN resulting in a small decrease in $T_c$. The suppression of
$T_c$ in high-DOS superconductors due to lattice disorder or interface effects
has also been reported previously \cite{murduck1987fabrication}. The high
values of $\lambda$ (approaching 2) place these NbN-based systems in the
strong-coupling regime, a phenomenon experimentally observed in similar
superconductors \cite{chen1971electron}.  Furthermore, the observed reduction
is consistent with experimental findings where minor interface imperfections
can slightly degrade superconductivity in heterostructures
\cite{zhang2013improvement}. In order to address the high value of $\lambda$ (1.75) for the $\delta$-NbN/AlN superlattice, the decoupling of smearing helps us get its value to be 0.86 and $T_c$= 10.83 K for $\sigma_{el}$=0.02 Ry. The decrease in $T_c$ (from 22.22 K to 10.83 K) results from a decrease in $N(E_F)$, causing EPC to become small \cite{koretsune2017efficient}.

We emphasize that superconductivity in these heterostructures is 
investigated within conventional EPC theory \cite{giustino2017electron}. While van Hove singularities 
associated with flat bands in other low-dimensional systems can occasionally 
trigger correlation-driven, non-phonon pairing mechanisms depending on 
filling and interactions \cite{markiewicz1997survey, leykam2018artificial, 
classen2025high}, our calculations show that standard lattice vibrations are 
sufficient to account for the estimated $T_c$ values without invoking 
non-local electronic correlations.

A critical question concerns the fate of the spontaneous polarization of AlN within these metallic superlattices. 
To quantify the polarization, we analyzed the atomic distortions in the 
relaxed structures (see Fig.\,\ref{fig:Superlattices}). 
We find that the internal parameter $u$ of the AlN layers remains nearly identical to the bulk value ($\approx 0.38$), 
whereas the $c$ lattice parameter of these layers is compressed by approximately 10\% relative to the bulk. 
Consequently, while the local dipole moments are reduced by roughly 10\% due to this compression, 
the macroscopic polarization inside the AlN layer remains effectively unchanged, as the reduction in dipole moment 
is compensated by the proportionate decrease in the unit cell volume. This result aligns well with the 
bulk wurtzite value ($P \approx 1.35$ C/m$^2$) \cite{karch1997ab,
zoroddu2001first, dreyer2016correct}, 
yet stands in stark contrast to our calculations for a freestanding 2D AlN layer. 
In the 2D limit, uncompensated depolarization fields force the layer to relax into a flat, non-polar structure ($u \approx 0.49$); 
in the superlattice, however, the polar structure is robustly preserved.

Despite the dimensional confinement, the system therefore maintains its non-centrosymmetric nature. 
The stability of this polar distortion in a metallic environment is 
governed by the screening provided by the TiN and NbN layers. 
This highlights the essence of the polar metal mechanism: free carriers in the TiN or NbN layers 
undergo charge redistribution that effectively (i) suppresses the depolarization field within the polar AlN layer, 
thereby preventing the flattening observed in the 2D limit, and (ii) generates an opposing electronic polarization within the conducting layer to screen the internal electric field \cite{pavlenko2003microscopic,pavlenko2004modulation}. 
Thus, the polar lattice distortion is not only retained but 
coexists with the metallic state, actively shaping the superconducting properties.

From a structural perspective, the alternating metallic and 
insulating blocks, which are atomically thin, create a structural geometry 
analogous to an intrinsic Josephson junction array. If the vertical coherence length and out-of-plane 
transport properties confirm proper decoupling across the AlN layers,  
which require further transport modeling in the vertical direction, 
such uniform, atomically thin configurations could provide a material 
platform for future investigations into low-dimensional superconducting 
electronics \cite{wang2008tunable, Savel'ev_2010, lei2014optimizing, wang2025thz}.

In summary, we have demonstrated that TiN/AlN and $\delta$-NbN/AlN superlattices host a robust superconducting state mediated by
interface-engineered phonons and enhanced by flat-band physics. These materials
represent a realization of superconducting polar metals, where polarity and
metallicity cooperate to define the ground state, effectively establishing a blueprint for the design of polar superconductors.
 We hope this study will
stimulate further investigation into these fascinating quantum structures
\cite{pavlenko2004modulation, dang2021all, balasubramanian2020epitaxial,
kobayashi2022crystal, kobayashi2022epitaxial}, opening new avenues for
combining electric polarization with superconducting quantum technologies.

\section*{Data Availability}
The data that support the findings of this study are available from the corresponding author upon reasonable request.

\section*{Acknowledgements}
This work was supported by the Texas Center for Superconductivity at the University of Houston (TcSUH). The authors acknowledge the computing resources provided by the HPE Data Science Institute and the Center for Advanced Computing and Data Systems at the University of Houston. The authors are grateful to Dr. Hongyan Lu for the helpful discussion. L.B. thanks the U.S. Department of Energy (DOE), Office of Science, Basic Energy Sciences (BES) under Award No. DE-SC0025479. The research was also funded in part by The Robert A. Welch Foundation under Grant No. L-E-0001-19921203.

\bibliography{biblio-Huma.bib}
 
\end{document}